\documentclass[twocolumn]{aastex631}

\usepackage{lineno}

\usepackage{graphicx}
\accepted{\today}
\usepackage{stackengine}
\stackMath
\graphicspath{{./}{figures/}}

\begin{document}

\title{Illuminating the Dark Side of Cosmic Star Formation \\ II. A second date with RS-NIRdark galaxies in COSMOS\footnote{Released on XXX}}

\correspondingauthor{Meriem Behiri}
\email{mbehiri@sissa.it}
\author[0000-0002-6444-8547]{Meriem Behiri}

\affiliation{SISSA, Via Bonomea 265, I-34136 Trieste, Italy}
\affiliation{University of Bologna - Department of Physics and Astronomy “Augusto Righi” (DIFA), Via Gobetti 93/2, I-40129 Bologna, Italy}
\affiliation{INAF/IRA, Istituto di Radioastronomia, Via Gobetti 101, 40129, Bologna, Italy}
\author[0000-0003-4352-2063]{Margherita Talia}
\affiliation{University of Bologna - Department of Physics and Astronomy “Augusto Righi” (DIFA), Via Gobetti 93/2, I-40129 Bologna, Italy}
\affiliation{INAF- Osservatorio di Astrofisica e Scienza dello Spazio, Via Gobetti 93/3, I-40129, Bologna, Italy}
\author[0000-0002-4409-5633]{Andrea Cimatti}
\affiliation{University of Bologna - Department of Physics and Astronomy “Augusto Righi” (DIFA), Via Gobetti 93/2, I-40129 Bologna, Italy}
\affiliation{INAF-Osservatorio Astrofisico di Arcetri, Largo E. Fermi 5, I-50125, Firenze, Italy}
\author[0000-0002-4882-1735]{Andrea Lapi}
\affiliation{SISSA, Via Bonomea 265, I-34136 Trieste, Italy}
\affiliation{INAF/IRA, Istituto di Radioastronomia, Via Gobetti 101, 40129, Bologna, Italy}
\affiliation{INFN - Sezione di Trieste, via Valerio 2, Trieste 34127, Italy}
\affiliation{IFPU - Institute for fundamental physics of the Universe, Via Beirut 2, 34014 Trieste, Italy}
\author[0000-0002-0375-8330]{Marcella Massardi}
\affiliation{INAF/IRA, Istituto di Radioastronomia, Via Gobetti 101, 40129, Bologna, Italy}
\affiliation{INAF, Istituto di Radioastronomia - Italian ARC, Via Piero Gobetti 101, I-40129 Bologna, Italy}
\affiliation{SISSA, Via Bonomea 265, I-34136 Trieste, Italy}
\author[0000-0002-0200-2857]{Andrea Enia}
\affiliation{University of Bologna - Department of Physics and Astronomy “Augusto Righi” (DIFA), Via Gobetti 93/2, I-40129 Bologna, Italy}
\affiliation{INAF- Osservatorio di Astrofisica e Scienza dello Spazio, Via Gobetti 93/3, I-40129, Bologna, Italy}
\author[0000-0002-8853-9611]{Cristian Vignali}
\affiliation{University of Bologna - Department of Physics and Astronomy “Augusto Righi” (DIFA), Via Gobetti 93/2, I-40129 Bologna, Italy}
\affiliation{INAF- Osservatorio di Astrofisica e Scienza dello Spazio, Via Gobetti 93/3, I-40129, Bologna, Italy}

\author[0000-0002-3915-2015]{Matthieu Bethermin}
\affiliation{Aix Marseille Univ, CNRS, LAM, Laboratoire d’Astrophysique de Marseille, Marseille, France}
\affiliation{Universit\'e de Strasbourg, CNRS, Observatoire astronomique de Strasbourg, UMR 7550, 67000 Strasbourg, France}
\author[0000-0002-9382-9832]{Andreas Faisst}
\affiliation{IPAC, M/C 314-6, California Institute of Technology, 1200 East California Boulevard, Pasadena, CA 91125, USA}
\author[0000-0002-8008-9871]{Fabrizio Gentile}
\affiliation{University of Bologna - Department of Physics and Astronomy “Augusto Righi” (DIFA), Via Gobetti 93/2, I-40129 Bologna, Italy}
\affiliation{INAF- Osservatorio di Astrofisica e Scienza dello Spazio, Via Gobetti 93/3, I-40129, Bologna, Italy}

\author[0000-0002-1847-4496]{Marika Giulietti}
\affiliation{SISSA, Via Bonomea 265, I-34136 Trieste, Italy}
\affiliation{INAF- Osservatorio di Astrofisica e Scienza dello Spazio, Via Gobetti 93/3, I-40129, Bologna, Italy}

\author[0000-0002-5836-4056]{Carlotta Gruppioni}
\affiliation{INAF- Osservatorio di Astrofisica e Scienza dello Spazio, Via Gobetti 93/3, I-40129, Bologna, Italy}
\author[0000-0002-7412-647X]{Francesca Pozzi}
\affiliation{University of Bologna - Department of Physics and Astronomy “Augusto Righi” (DIFA), Via Gobetti 93/2, I-40129 Bologna, Italy}
\affiliation{INAF- Osservatorio di Astrofisica e Scienza dello Spazio, Via Gobetti 93/3, I-40129, Bologna, Italy}

\author[0000-0002-3893-8614]{Vernesa Smolçić}
\affiliation{Department of Physics, Faculty of Science, University of Zagreb Bijenićka cesta 32, 10000 Zagreb, Croatia}
\author[0000-0002-2318-301X]{Gianni Zamorani}
\affiliation{INAF- Osservatorio di Astrofisica e Scienza dello Spazio, Via Gobetti 93/3, I-40129, Bologna, Italy}

\begin{abstract}
About 12 billion years ago, the Universe was first experiencing light again after the dark ages, and galaxies filled the environment with stars, metals and dust. How efficient was this process? How fast did these primordial galaxies form stars and dust? We can answer these questions by tracing the Star Formation Rate Density (SFRD) back to its widely unknown high redshift tail, traditionally observed in the Near-InfraRed (NIR), Optical and UV bands. Thus, the objects with a high amount of dust were missing. We aim to fill this knowledge gap by studying Radio Selected NIR-dark (\textit{RS-NIRdark}) sources, i.e. sources not having a counterpart at UV-to-NIR wavelengths. We widen the sample by Talia et al. (2021) from 197 to 272 objects in the COSMic evolution Survey (COSMOS) field, including also photometrically contaminated sources, previously excluded. Another important step forward consists in the visual inspection of each source in the bands from u* to MIPS-24$\mu$m. According to their "environment" in the different bands, we are able to highlight different cases of study and calibrate an appropriate photometric procedure for the objects affected by confusion issues. {We estimate that the contribution of RS-NIRdark to the Cosmic SFRD at 3$<$z$<$5 is $\sim$10--25$\%$} of that based on UV-selected galaxies. 
 
\end{abstract}

\defcitealias{talia21}{T21}
\section{Introduction} \label{sec:intro}

Understanding the cosmic star formation history is fundamental to improve our knowledge of galaxy formation and evolution, the production of heavy elements and the re-ionization process through cosmic time. In the ‘90s \cite{lilly}, and \cite{madau1996} built the Lilly-Madau plot with Hubble Space Telescope (HST) data to study the evolution of the Star Formation Rate (SFR) per unit of cosmic volume (i.e. the Star Formation Rate Density, SFRD) as a function of redshift. Thanks to the critical technological improvements, scientists have added more observational points to this plot, improving the constraints on the SFRD and pushing this study toward higher redshifts. \cite{mdauadickinson} show an evolution of the SFRD $\propto(1+z)^{1.7}$ up to z $\sim$ 2, the so-called cosmic-noon, while between z $\sim$ 3 and z $\sim$ 8 the SFRD decreases $\propto (1+z)^{-2.9}$. The evolution of the SFRD through cosmic time is now well constrained until z $\sim$ 3.
However, the possible decline of the SFRD after z $\sim$ 3 is still an open question \citep{casey2014,magnelli,zavala}, and some works argue that it might flatten \citep{2013MNRAS.436.2875G,rowanro,Novak2017,gruppioni2020}. 

Many high-redshift studies of the SFRD mainly rely on rest-frame UV/optically bright sources, mostly from Lyman-break galaxies (LBGs). The lack of dusty galaxies in the census introduces biases in the estimations of the SFRD since the LBGs represent only a fraction of the whole galaxies population, and the computation of the SFR for these sources strongly depends on the dust corrections adopted. We have recently witnessed a surge of candidates at very high-\textit{z} discovered by the \textit{James Webb Space Telescope} (JWST) (e.g. \citealt{naidu+22, Castellano+22, Adams+22, Finkelstein+22, Atek+22, Yan+22}); however, these measurements are mostly in the UV rest-frame, therefore a dust-unbiased view of the high-\textit{z} Universe is still fundamental to reconstruct the SFRD. 

On the one hand, \cite{rowanro}, using IR data from the {Herschel} Telescope, found that, at z$>$3, the contribution of the dusty galaxies to the SFRD could be about ten times that due to UV-bright sources. On the other hand, studying high-redshift galaxies in the infrared and submillimeter bands has always been technically tricky since observations are deeply affected by confusion problems \citep{lutz2011,magnelli2013}. Moreover, the faintness and the lack of counterparts in the UV$/$optical$/$NIR bands made identifying dusty galaxies at z $>$ 3 even more challenging. The first deep surveys in the (sub)millimetric bands opened the path to the studies on these galaxies at high redshift, taking advantage of the negative \textit{k}-correction in this part of the spectrum \citep{smail,hughes98}. \cite{smail} observed for the first time dusty submillimeter galaxies with the Submillimeter Common-User Bolometer Array (SCUBA) instrument of the James Clerk Maxwell Telescope (JCMT). These are the well-known SubMillimiter Galaxies (SMGs), i.e. galaxies with S$_{850\mu m}>$3 mJy \citep{smg}, and many of them resulted in being at high redshift. The South Pole Telescope (SPT) has given a notable contribution to this field, allowing the identification of $\sim$40 candidate Dusty Star-Forming Galaxies (DSFGs) at 5$<$z$<$6 \citep{strandet2016}, together with the Atacama Large Millimetre Array (ALMA). ALMA enabled the detection of many of these objects by tracing dust in the continuum (e.g., \citealt{scoville2014,scoville2016,scoville2017,brisbin2017,gonzalez2017,franco2018,dudze2020,smail2021}).

In fact, prior to the arrival of ALMA data, a thorough comprehension of this kind of objects was not conceivable (e.g., \citealt{Dannerbauer2004,Frayer2004}). For instance, thanks to the high spatial and spectral resolution of ALMA, \cite{walter12} spectroscopically confirmed the redshift 
of the starburst galaxy HDF850.1, detected with SCUBA by \cite{hughes98}. As more data became available, the idea that most high redshift dust-obscured galaxies were rare, incredibly star-forming or Active Galactic Nuclei (AGN), has been revisited, since normal galaxies started to emerge from the dust (e.g., \citealt{Simpson2014,Simpson2017,simpson2020,franco2018,wang2019}), and they could be more common than expected. These studies confirmed the hypothesis that optical/NIR studies missed a significant fraction of objects in the high redshift galaxies census.
Nevertheless, also (sub)mm and far-infrared (FIR) surveys have some drawbacks. These bands sample the dust grey body: since the dust temperature might correlate with infrared luminosity and redshift \citep{bethermin2015,faisst2017,schrei2018}, this could generate a  selection bias. Furthermore, ALMA deep fields are still not wide enough ($\sim$ 4.5-10 arcmin$^2$ \citealt{walter2016,Aravena_2016,dunlop17,franco2018,hatsukade2018}) to provide a complete view of these objects. In the end, since every selection method affects the results differently, combining different approaches and exploring new strategies is vital.

Primordial dusty galaxies are now ubiquitous in theoretical models of galaxy formation as possible progenitors of massive red and dead galaxies \citep{Cimatti2004,Cimatti2008,Cimatti2008nature,casey2014,ManLap2016,Mancuso2016,lapi18}. Moreover,  it has been recently suggested that these galaxies could be the most likely hosts of high redshift gravitational waves that the Laser Interferometer Space Antenna (LISA) and the Einstein Telescope (ET) will study in the next future \citep{boco}. 
Our work fits into this context, aiming to provide a deeper understanding of the role of obscured galaxies in galaxy formation and evolution.
We propose an alternative approach to the traditional ones used to select obscured galaxies (e.g. (sub)mm and FIR selections), on the path of \cite{talia21}, hereafter \citetalias{talia21}. They selected radio sources in the COSMic evolution Survey (COSMOS) field having no (or extremely faint) NIR counterparts (see Section \ref{sec:sample}). We call these sources Radio Selected NIR-dark (RS-NIRdark) galaxies. \citep[see also][]{chap2001,chap2002,chap2004,Enia} The K$_{S_{\rm lim}}$ adopted is $\sim$ 24.7, which is the Ks limiting flux density {at 3$\sigma$ }of the COSMOS2015 catalogue \citep{laigle2016}.

Radio-selected sources give many advantages: radio interferometers reach high sensitivity and resolution, plus dust temperature bias does not affect the selection. The most significant bias possibly affecting radio selection is the possible presence of AGN, but we can deal with this problem thanks to multi-wavelength ancillary data. Galaxies in our study are, by the selection, not (or barely) detected in optical and NIR bands; therefore, we need mid-infrared (MIR), FIR and millimetric (mm) data in order to reconstruct their Spectral Energy Distributions (SEDs). Thus, the availability of multi-wavelength catalogues is fundamental to study these sources. {Another drawback of the radio selection is the effect of the positive $k$-correction, which makes more challenging going towards higher redshifts. }For these reasons, we selected our sources from the VLA-COSMOS 3 GHz Large Project \citep{smolcic2017} catalogue, a deep radio survey in the COSMOS field that benefits from many ancillary data.

This paper is organized as follows. In Section \ref{sec:sample} we explain the sample selection process and the features of the different (sub)samples. In Section \ref{sec:median} we describe the median properties of the sample. In Section \ref{sec:indpho} we present the study of the individual photometry of our objects, with a particular focus on the new deblending procedure and on the presence of 500 $\mu$m risers. In Section \ref{sec:indsed} we discuss the results from the SED-fitting of the individual sources. In Section \ref{sec:sfrd} we derive the SFRD and the number density for our sample. 

Throughout this paper we adopt a \cite{imf} initial mass function (IMF) and we assume a cosmology with matter density $\Omega_M=0.3$, dark energy density $\Omega_{\Lambda}=0.7$ and Hubble parameter $H_0=70$ km$^{-1}$Mpc$^{-1}$. Magnitudes are given in the AB system.

\section{Sample Selection} \label{sec:sample}

\begin{figure*}[t]
\centering
\includegraphics[width= 1.\textwidth]{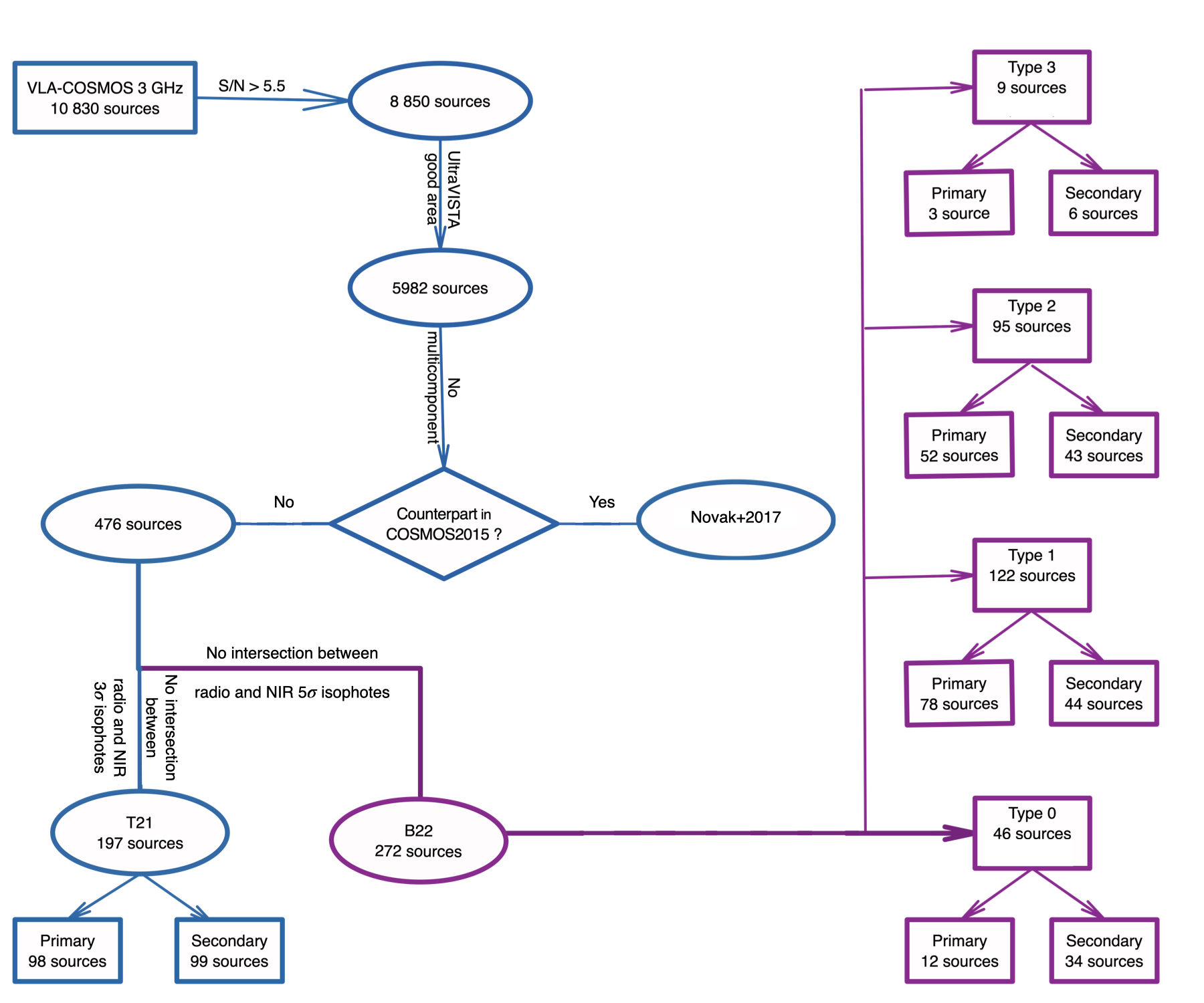}
\caption{Flow chart representing the selection process of the sample. The dark blue part of the chart indicates those steps made by \citetalias{talia21}, while the purple contours are used to highlight the ones made in this work.  }\label{fig:flow}\end{figure*}

\begin{figure*}[t] 
\centering

\includegraphics[width=0.9\textwidth]{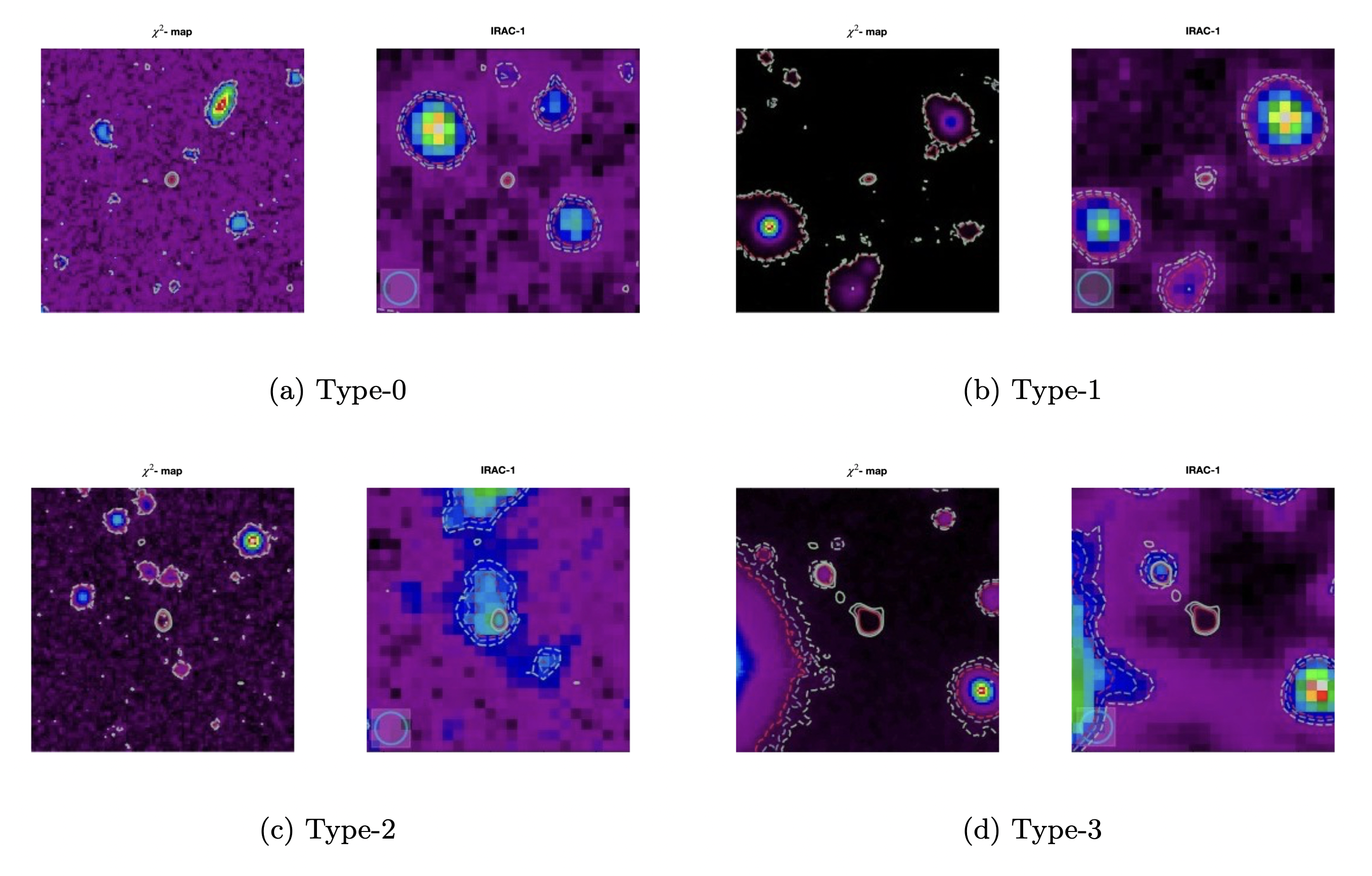} 
\caption{In this figure we show the galaxies belonging to each of the four sub-samples. We present examples for each group, show their $\chi^2$ and IRAC-1 cutouts, to which we superimposed the radio (solid lines) and the $\chi^2$ and IRAC-1 (dashed lines) contours at 3, 4 and 5 $\sigma$ (crimson). \textit{From the upper-left:} {(a)} \textbf{Type-0}: galaxies with no counterparts up to MIPS-24$\mu$m; (b) \textbf{Type-1}: isolated galaxies in all bands and with a counterpart in the MIR filters; (c)\textbf{Type-2}: galaxies having a counterpart and a nearby contaminant in one or more bands; (d) \textbf{Type-3}: like Type-0s, but close to another radio-source with a counterpart in one or more bands.}\label{fig:selec}
\end{figure*}

We selected the galaxies studied in this work from the VLA-COSMOS 3 GHz Large Project \citep{smolcic2017} catalogue, based on a survey covering 2.6 deg$^2$ in the COSMOS field in 348 hours of observations with the Karl G. Jansky Very Large Array (VLA) interferometer at 3 GHz. Thanks to its limiting flux density of {12.6} $\mu$Jy beam$^{-1}$ at 5.5 $\sigma$, this data set is one of the deepest ever obtained, suitable for estimating the contribution to the cosmic SFRD of galaxies {at least up to z$\sim$5}, as shown in \cite{Novak2017} and \citetalias{talia21}. 
We summarise the selection process in Figure \ref{fig:flow}.

This work inherits the first steps of the selection process from \citetalias{talia21}. They selected 8850 sources detected with an S/N $>$ 5.5 from the VLA-COSMOS 3 GHz \citep{smolcic2017}. They choose this S/N threshold to minimize the fraction of spurious sources. Subsequently, \citetalias{talia21} excluded all the sources within the bad areas of the  VISTA Ultra-Deep (UltraVISTA) field \citep{laigle2016}, reducing the effective area reduces to 1.38 deg$^2$, and discarded all the multi-component objects. They cross-correlated the resulting catalogue of 5982 sources with the COSMOS2015 catalogue \citep{laigle2016} within a search radius of 0.8'' \citep{smolcic2017}, finding 476 sources without a counterpart. These are the RS-NIRdark (according to the Ks$_{lim}$ of COSMOS2015, i.e. $\sim24.7$ mag) galaxies which constitute the starting point of this work.
As an additional step, \citetalias{talia21} narrowed down their selection to the 197 sources whose 3$\sigma$  isophotes in the 3 GHz and $\chi^2$ maps\footnote{The NIR detection map used to build the COSMOS2015 catalog is a $\chi^2$  composite image of the $z++$, Y, J, H and K$_s$ maps, where each pixel has a certain probability of belonging or not to the background. Knowing the probability distribution of the background pixels (a $\chi^2$ distribution), it is possible to use the data to obtain the distribution of the pixels dominated by the source emission and to determine the best threshold to identify the background pixels \citep{szalay99}.}, respectively, do not intersect. 
In our work we relax this constraint and include all the sources whose 5$\sigma$ 3 GHz isophote does not intersect the 5$\sigma$ NIR isophote, totaling 272 objects. Thus, \citetalias{talia21} selection is stricter than ours and includes only the most isolated galaxies, while we extend the analysis to more contaminated sources, which requires the use of more refined photometric techniques.
The analysis of the remaining 204 sources, which are heavily contaminated by nearby NIR-bright galaxies,  will be presented in a forthcoming paper (Gentile et al. in prep.).

We split the 272 objects into 4 sub-samples based on visual inspection of the NIR band maps from {the second} UltraVISTA Data Release (DR2) \citep{cracken} and InfraRed Array Camera (IRAC) and Multiband Imaging Photometer for Spitzer (MIPS) maps {from} \textit{Spitzer}\footnote{We used the UltraVISTA DR2 maps for the selection for consistency with \citetalias{talia21}. The IRAC maps are the v2.0 mosaics of the
Spitzer Large Area Survey with Hyper-Suprime-Cam
\citep[SPLASH;][]{steinhardt2014} \url{https://splash.caltech.edu/}.}:
\begin{itemize}
\item \textbf{Type-0}: Isolated galaxies, with no counterparts up to MIPS-24$\mu$m.
\item \textbf{Type-1}: Isolated galaxies both in the $\chi^2$ and IRAC-1 maps, having a counterpart in
one or more bands.
\item \textbf{Type-2}: Galaxies with a counterpart in one or more bands which are also contaminated by nearby sources. Objects belonging to this class might show a distinct brightness peak, with respect to the contaminant, in all the bands, or they might present a distinct peak in the optical and NIR bands but be unresolved in the MIR images.
 \item \textbf{Type-3}: Like Type-0s, but close to another radio source with a counterpart in one or
more bands.
\end{itemize}
We display examples and explanations of the different sub-samples in Figure \ref{fig:selec}.

We stress that  \citetalias{talia21} selected the most  \emph{isolated} galaxies  based  on the inspection of the radio and NIR maps only. This means that a fraction of \citetalias{talia21} sources presented indeed some, albeit low, degree of contamination by a nearby source in the MIR bands.
In this work, thanks to the  visual inspection of the maps at all wavelengths, we have not only enlarged the sample but also improved the overall photometry with respect to the previous study.

We matched our catalog also to the \textit{Heschel} Super-Deblended Catalog by \cite{Jin} for the FIR bands and, where available, the Automated Mining of the ALMA Archive in the COSMOS Field (A3COSMOS; \citealt{liu2019}) for the (sub)mm bands. Finally, we introduce an additional classification based on the FIR/(sub)mm coverage of our sources:
\begin{itemize}

 \item \textbf{Primary Sample}: sources having at least one detection{above 3$\sigma$} in one of the FIR/(sub)mm bands. It accounts {for} 145 sources.
 \item \textbf{Secondary Sample}: sources with no detections above 3$\sigma$ in any FIR bands. It accounts {for} {127} sources.
\end{itemize}

Since we want to estimate the contribution to the cosmic SFRD of the RS-NIRdark sources, we only consider the former since it leads to better constraints on redshift and the star formation rate, which we compute from the IR luminosity.
\section{Median properties of the sample} \label{sec:median}

\subsection{Median Photometry}
\label{sec:medfot}
We start by considering the {average properties} of the sample as a whole through statistical analysis, by performing a median stacking. In this procedure, we exclude Type-3 sources, 
because these might be jets belonging to the nearby radio source rather than galaxies themselves. {After the individual analysis of the sources (Section \ref{sec:indsed}), we have considered the possibility of performing a tomographic stacking, dividing the sample into redshift bins. Nevertheless, the number of sources in the high redshift bins would be exiguous, thus, the median stacking would not have the desired effect.}

We build the median maps from u* to MIR by stacking individual maps at the VLA 3 GHz coordinates of the radio-source. In particular, we used: Y, J, H and K$_s$ maps from UltraVISTA DR4\footnote{This is the same UltraVISTA data release used in the recent COSMOS2020 catalogue \citep{cosmos2020}.}, IRAC maps from SPLASH\footnote{The IRAC maps used in the COSMOS2020 catalog are instead those from \citet{moneti2022}: the difference is minimal and only concerns channels 1 and 2, as shown in Figure 3 of \cite{cosmos2020}.},  MIPS maps from the Spitzer COSMOS (SCOSMOS) program. In the optical regime we used the same maps as in \citet[][; see their Table 1 for a summary]{laigle2016}. 
From the stacked maps, we extract the median fluxes using Source Extractor \citep[SExtractor;][]{SE} for the bands from H to MIPS-24$\mu$m, while from u* to J, where the stacked flux is at best marginally detected, we choose the Aperture Photometry Tool \citep[APT;][]{apt} to execute this analysis. 

In the FIR regime we compute the median fluxes by performing survival analysis \citep{survival}, which takes into account the presence of upper limits,  on the Herschel counterparts from the aforementioned Super-Deblended catalogue by \cite{Jin}. {The choice of not performing the image stacking for these bands is due to the coarse resolution of the FIR observations and the ensuing difficult extraction of the sources from the maps.}

\subsubsection{Type-0 }
\begin{figure*}[t]
 \centering
 \includegraphics[width= 0.9\textwidth]{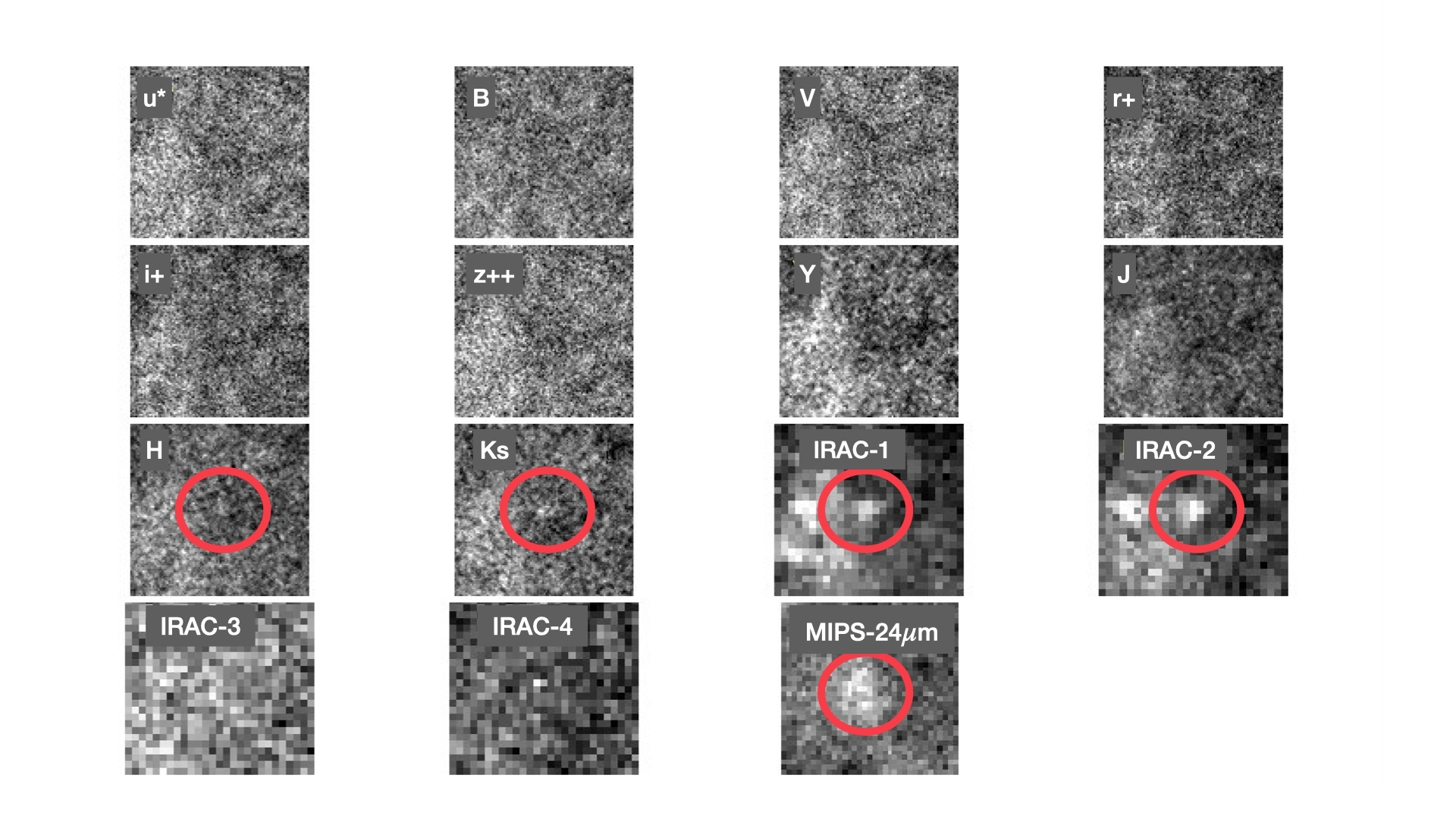} 
 \caption{Median stacked images of the Type-0 sub-sample in the u* to MIPS-24$\mu$m bands. The red circles indicate the bands where we find a stacked flux with S/N$>$2.}
 \label{fig:fot0}
\end{figure*}

We perform the same procedure also for the Type-0 sub-sample alone and we show the stacked images in Figure \ref{fig:fot0}: although these sources do not have any counterpart up to MIPS-24$\mu$m map, the median stacking reveals in H, K$_s$, IRAC-1, IRAC-2 and MIPS-24$\mu$m an emission at S/N$>$2. 
Many of the individual sources belonging to this category have extremely bright radio fluxes, up to 5.2$\pm$0.3 $\mu$Jy, and show peculiar radio morphologies; furthermore, 12 over 34 of these objects belong to the Primary sample, i.e. they have a counterpart in at least one of the FIR/(sub)mm bands. These sources could be heavily obscured galaxies and high redshift candidates. However,  SED-fitting is  inconclusive for these sources, as would be expected given the very poor photometric coverage currently available. Only with more data we might be able to retrieve some information on this sub-sample, which is particularly suited for deeper  follow-up observations with ALMA and the JWST.

\begin{figure*}[t]
 \centering
 \includegraphics[width= 0.9\textwidth]{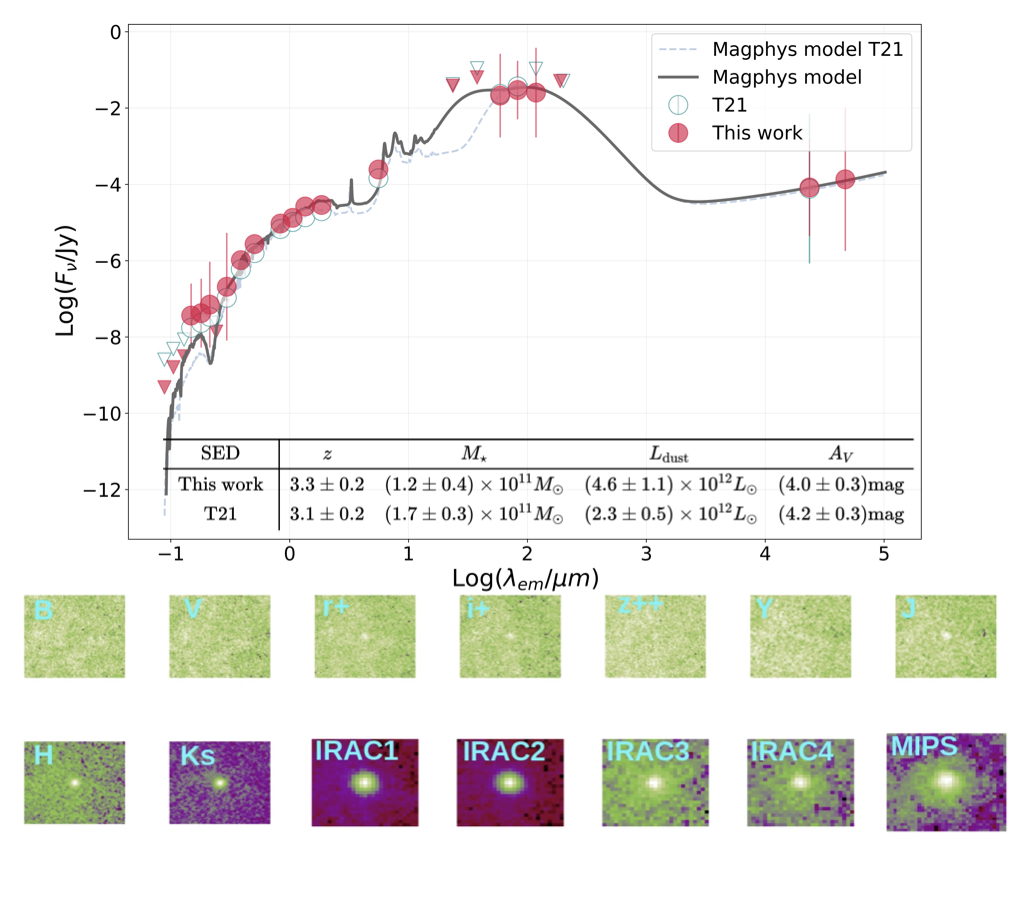} 
 \caption{\textit{Top}: Median SED of Type-0+1+2 galaxies. Crimson and empty teal points and arrows represent, respectively, this work and \citetalias{talia21} detections and 1$\sigma$ upper limits. The best-fit model derived for our stack is the dark gray line {for this work, while for \citetalias{talia21} it is represented by the light gray dashed line}. {The SED fitting is performed on the median stacked photometry from the UV to the mid-IR bands and on the median fluxes from the \emph{Super-deblended} catalog \citep{Jin} in the FIR regime, as explained in the text.} The table shows some of the {properties} from the SED-fitting, i.e. photometric redshift, stellar mass, dust luminosity, dust temperature and extinction, for both \citetalias{talia21} and this work. The errors are computed as {the semi-interval between the 16th and the 84th percentile of the probability distribution functions of each parameter}. The quoted error on the redshift includes also a component from bootstrap analysis on the stacked photometry. \textit{Bottom}: median stacking of the Type-0+1+2 sub-samples in the u* to MIPS-24$\mu$m bands. }
 \label{fig:fot012}
\end{figure*}

\subsection{Median SED-fitting}
We derive the median properties of the  Type-0+1+2 sources by performing SED-fitting on the median stacked photometric data, from u* to 1.4 GHz, described in the previous section. To this purpose, we use the \textsc{MAGPHYS+photo-z} code \citep{2015dacunha,magphys}. 

\textsc{MAGPHYS} is a SED-fitting code based on energy balance. {It} derives the {physical} properties of a galaxy through the {comparison of} the observed photometry {to} a large number of models: for each of the possible parameters, the code builds a likelihood distribution to choose the one that best fits the observations. \textsc{MAGPHYS} can model a large variety of galaxies, as shown by \cite{magphys}. The photo-z version of \textsc{MAGPHYS}, \textsc{MAGPHYS+photo-z}, allows us to compute the redshift along with the other properties of the galaxy. This feature brings numerous advantages: fitting all the properties simultaneously means having more robust uncertainty parameters.
One of the main reasons for choosing \textsc{MAGPHYS} in this work is that the code uses all the photometric points up to the radio band, which is particularly suited to a sample like ours where the coverage in the optical and NIR bands is very scarce.

For the stellar component, \textsc{MAGPHYS}  uses the Simple Stellar Populations (SSPs) from \cite{ssp}, assuming a \cite{imf} IMF, with delayed exponentially declining star-formation histories combined with random bursts. Dust attenuation is taken into account as described by \cite{charlot}, with the addition of the 2175 \AA feature \citep{Battisti_2020}, while dust emission modelling takes into account both the components due to stellar birth clouds and the one due to the ambient ISM \citep{2008dacunha}. The radio component model relies on the hypothesis that the IR-radio correlation parameter $q_{TIR}=2.34$. For the metallicity, \textsc{MAGPHYS}  uses a prior uniformly distributed between 0.2 and 2.0 solar metallicities. For further details on the code, see \cite{Battisti_2020}.

From the SED-fitting procedure {on the stacked photometry }of Type-0+1+2 sub-sample, we obtain the {median} properties of the galaxies in our catalogue. 

The resulting best-fit SED is shown in Figure \ref{fig:fot012}. The median properties of the sample are consistent with those of Ultra-Luminous Infrared Galaxies (ULIRGs), since the median infrared luminosity is as high as $L_{IR}=(8.48\pm0.05)\times10^{12}L_\odot$, corresponding to a SFR$\sim100 M_{\odot}$yr$^{-1}$. This value  is broadly consistent, though {slightly }higher, with that by  \citetalias{talia21} ($L_{IR}=(2.3\pm0.5)\times10^{12}L_\odot$).
The median photometric redshift\footnote{The error on the redshift is the quadratic sum of the output from \texttt{MAGPHYS} and of the result of bootstrap analysis on the stacked photometry.} is $z_{med}=3.3\pm0.2$, which is also consistent with \citetalias{talia21} ($z_{med}=3.1\pm0.2$) and suggests the presence of a relevant number of high-redshift ($z>4.5$) galaxies in our sample.

In order to investigate the possible presence of obscured AGN in our sample, we extend our analysis toward the high-energy side of the spectrum. {T}hanks to the CSTACK  online tool, we stacked X-ray data  in the  soft band ($[0.5-2]$KeV)  from the Chandra-LEGACY survey \citep{civano}. {After the exclusion of X-ray point sources, }assuming $N_H=10^{22}$cm$^{-2}$ and a photon index $\Gamma = 1.8$, the resulting rest-frame luminosity is $L_{2-10 KeV}\sim10^{42}$ ergs$^{-1}$, which  suggests the absence of powerful AGN in our sample. In fact, this X-ray emission can be attributed to the star formation process alone, since it corresponds to a SFR $\sim 100M_{\odot}$yr$^{-1}$, according to the \cite{Kennicutt_2012} relation, in perfect {agreement }with the SFR computed through $L_{IR}$.

The small differences in the median results with respect to \citetalias{talia21} are mainly due both to the widening of the sample, which enabled us to have more constraining upper-limits, and to the fact that the new sources are overall more disturbed than the ones studied in the previous work. 

To test the representativeness of the results obtained with the stacking analysis, we also
compared the best-fit SED obtained from the the median stacked photometry of the primary sample alone to the median of the best-fit SEDs of the individual
galaxies described in the next section. 
From this test, it emerges that the two composite SEDs are almost coincident,  the main difference being that in the former the FIR peak is wider.
This is due to the redshift distribution of the sample sources, which artificially enlarges the bell in the Herschel region of the SED. These deviation,
however, do not have any dramatic effect on the redshift nor on the recovered median properties.

\section{Individual Photometry}
\label{sec:indpho}
We perform the analysis of the individual sources as well, where the visual inspection allows us to choose the best technique to process the four different types. This  is a step forward with respect to \citetalias{talia21}: in this work, in fact, we present brand new photometry for the sources with confusion issues in the IRAC bands that takes into account the {possible}  contamination {by nearby sources}.

\subsection{Optical and NIR}
\label{sec:optnir}
By selection, our sources are either not detected or extremely faint in the optical bands and, as such, in these bands the flux densities are given as upper limits. In particular, we set the threshold for considering whether the source is detected or not in a given band at S/N$=$3. 

In particular we use those reported by \cite{laigle2016}. 

As for the NIR bands, we took advantage of the UltraVISTA DR4 catalogue\footnote{\url{https://ultravista.org}}, deeper than the UltraVISTA DR2 on which the COSMOS2015, hence our selection, is based. 

We can distinguish two different cases.
First,  radio sources that have a counterpart with S/N $>$ 3 in the deeper images (128 objects). In this case we took the flux density and associated errors from the existing catalog. For the marginally detected sources we set as 1$\sigma$ upper limit the error on the flux. We note that most of these sources have therefore a counterpart also in COSMOS2020 \citep{cosmos2020}, the most recent version of the multi-wavelength photometric catalog in the COSMOS field, which is also based on the UltraVISTA DR4. We briefly discuss this point in Appendix \ref{appendix}.
We stress that, given that our selection was based on the lack of a counterpart in the COSMOS2015 catalog, we decided not to treat separately these sources.

Second, sources with no counterpart in the deeper images. In this case, we measure the 1$\sigma$ upper limit directly on the  maps within a fixed aperture of 2'' using the Python \textsc{Photutils} package \citep{phot}.

\subsection{MIR}
We extract the flux densities in the MIR bands from the IRAC-1,2,3 $\&$ 4 and MIPS-24$\mu$m maps from the v2.0 mosaics of SPLASH using SExtractor (SE; \citealt{SE}) {for all objects, except those belonging to Type-2 class (see Section \ref{sec:sample}), whose MIR photometry is affected by blending problems due to the scarce spatial resolution. For these sources we perform the photometry with specific methods, described in  Section \ref{sec:deb}.}
{For the objects not suffering from blending, we first make a blind run on the MIR maps with SE on all the sources. This procedure generates a catalogue that we match with our sources using a 1.7'' radius \citep{smolcic2017} to find the MIR flux densities. If the MIR counterpart has an S/N$<$3, we put an upper limit equal to the error on the extracted flux itself. If the source has no counterpart in the SE catalog, we measure the upper limits with \textsc{Photutils} with an aperture size of 3.6''.}
{\subsubsection{The deblending procedure}
\label{sec:deb}
Following a visual inspection, we isolate a sub-sample of 95 galaxies (Type-2; see Section \ref{sec:sample}) which suffer from contamination by one or more adjacent sources in the IRAC bands. 

For these galaxies we did not obtain  satisfying results using SE, even when forcing the aperture photometry at the radio position in double mode and then correcting for the aperture loss. Thus, we decide for a different approach to  extract the fluxes.

We use the Python package c \citep{lang2016}, which is particularly efficient in the deblending of sources similar to ours\footnote{In IRAC 4 and MIPS the deblending procedure failed because of their too coarse resolution, so we used the blended flux by SE in these bands.}, as already shown by \citet{Enia} \citep[see also][]{cosmos2020}}

\begin{figure}[h]
    \centering
    \includegraphics[width= 0.45\textwidth]{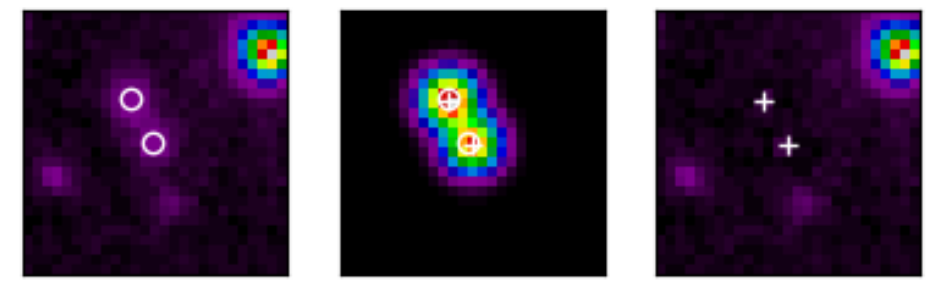}
    \caption{Example of the deblending procedure on an IRAC-1 map. The first image is the input map, with the input positions (white circles). The second image is the model produced by \textsc{Tractor}, where the white crosses indicate the best fit positions. The last image is the residual map.}
    \label{fig:tractor}
\end{figure}

Starting from defined astronomical sources with certain specified properties (i.e. position, flux and so on), the code gives estimates in the pixel space on what should be found in the observed image. The input map is the IRAC map centered on the radio coordinates of the RS-NIRdark source. Therefore, the IRAC map should show the MIR counterpart of this source. However, as we said before, sometimes this counterpart is contaminated by the flux from another or more sources. The contaminants are usually sources belonging to the COSMOS2015 catalogue. Thus, we give as prior positions to \textsc{Tractor} the radio and COSMOS2015 coordinates and an initial guess on the flux value of all the sources involved (target and contaminants). We set this initial guess as a fraction of the total blended flux extracted with SE. We summarize the input parameters in Table \ref{tab:tractor}, while in Figure \ref{fig:tractor} we show the results on an IRAC-1 map of a galaxy in our sample. 

\begin{table*}[ht]
\begin{center}

\begin{tabular}{|c|c|c|}
\hline
\textbf{Input parameter} &      \textbf{Description}     & \textbf{This work} \\
\hline
      \textbf{Map}         &     Blended map      & IRAC-1 27$\times$27 pixels$^2$ map \\

                \hline
\textbf{Positions} & {Contaminants positions
} &  {The COSMOS2015 position of the} \\
          &    and target position       &     contaminant $+$ the VLA-COSMOS 3GHz      \\
         &           &     position of the target      \\
           \hline
       \textbf{Fluxes}    &   Best guess fluxes for the target        &    SE blended flux/N;       \\
          &   and the contaminants        &     N$=$number of contaminants + 1      \\
          &           &      $=$ 2 if 1 contaminant    \\
    \hline
      \textbf{PSF}                 & FWHM of the IRAC band considered & 1.7''\\
          \hline
      \textbf{Per-pixel image noise}              &FWHM of the map & 0.1\\
                 \hline
              \textbf{dlnP}                  & Precision of the linearized least squared for the best fit & 10$^{-3}$\\
\hline
           \textbf{Profile}        & Gaussian with $\sigma =$ FWHM & Gaussian with $\sigma=$1.7 \\
\hline
\end{tabular}
\caption{Description of the input parameters for the deblending procedure with \textsc{Tractor}, and the specific values used in our work. In Figure 5 we show the input map, the output model and the residuals relative to the example described in this table.}
\label{tab:tractor}
\end{center}

\end{table*}
    
\textsc{Tractor} produces a mask containing the positions and the fluxes given as input. We let \textsc{Tractor} free to move inside the map to find the brightness peaks. Through a Markov Chain MonteCarlo, the code builds a best-fit model and creates a catalogue containing the deblended sources flux. 

We compared the residual maps obtained after subtracting the blended source+contaminant detection by SE with those obtained after subtracting the de-blended source and contaminant by \textsc{Tractor}: they are practically identical, meaning that the two methods recover the same total flux densities in the source+contaminant region. This constitutes additional proof in favor of the reliability of our results. With this method, we successfully deblended 46 of the 52 Type-2 primary sources. 

\subsection{FIR and (sub)mm}
We extend our photometric range to the FIR thanks to the \textit{Herschel} Super-Deblended Catalog by \cite{Jin}.{We find a counterpart at the 3$\sigma$ significance level in at least one \textit{Herschel} band for 139
galaxies, out of 272. For marginally detected sources we set as 1$\sigma$ upper limits the error on the flux, when available.}
We {also} perform a match within a radius of 0.8'' between our data and the  A3COSMOS catalog\footnote{Version v.20200310.}, which contains photometric data from the ALMA archive. We find counterparts at S/N$>$3 in at least one ALMA band for 34 galaxies. Given the extremely heterogeneous nature of the observations used to build the catalog, it is impossible to define upper limits. Therefore, when no counterpart is present in the A3COSMOS catalog in the ALMA bands, we exclude the ALMA point from the photometric SED that we give as input to MAGPHYS.

We recall that the 145 galaxies with at least one 3$\sigma$ detection in the FIR/(sub)mm range constitute our \emph{primary} sample (see Section \ref{sec:sample}), which we use as the basis of our SFRD computation.
We perform individual SED-fitting analysis for the primary sample only, for which we expect more reliable results because of the broader photometric coverage from MIR to FIR/(sub)mm.

\subsubsection{500 $\mu$m risers}
\begin{figure} 
\centering

\includegraphics[width=0.5\textwidth]{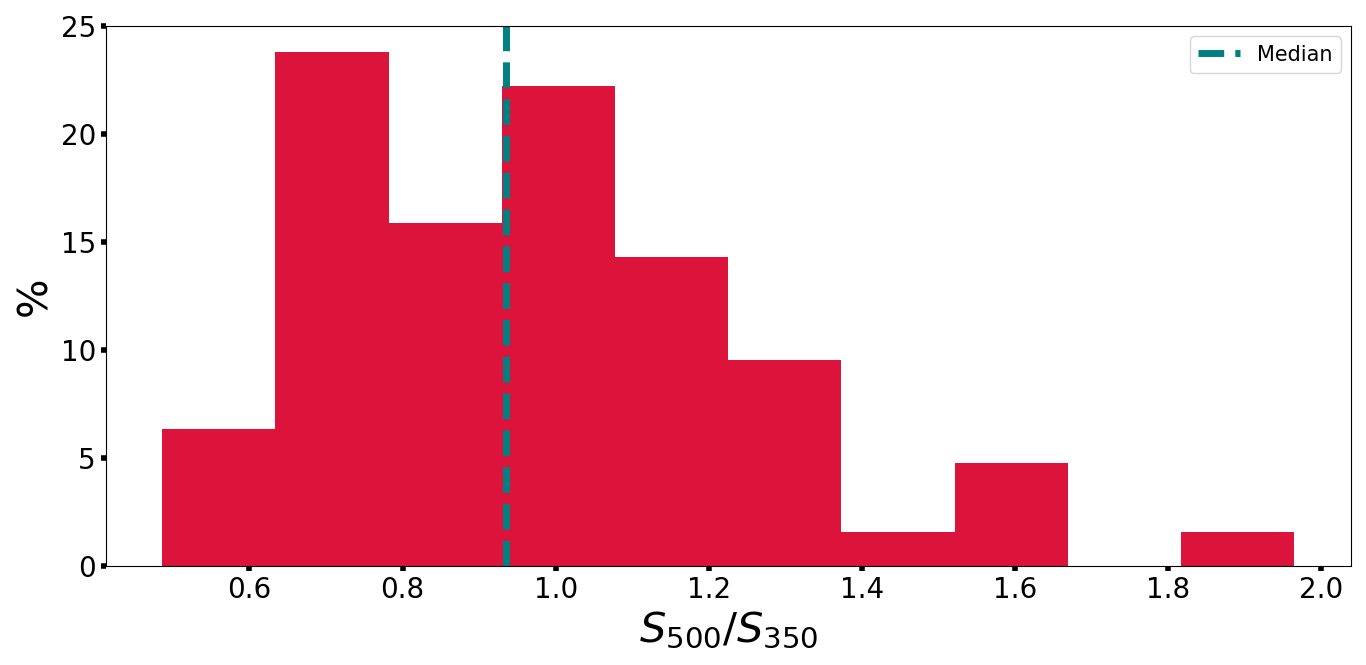}
\caption{Histogram showing the percentage of galaxies in our primary sample in a given S$_{500}$/F$_{530}$ bin. The dashed line indicates the median value.} \label{fig:histspire}
\end{figure}

\begin{figure*} 
\centering

\includegraphics[width=\textwidth]{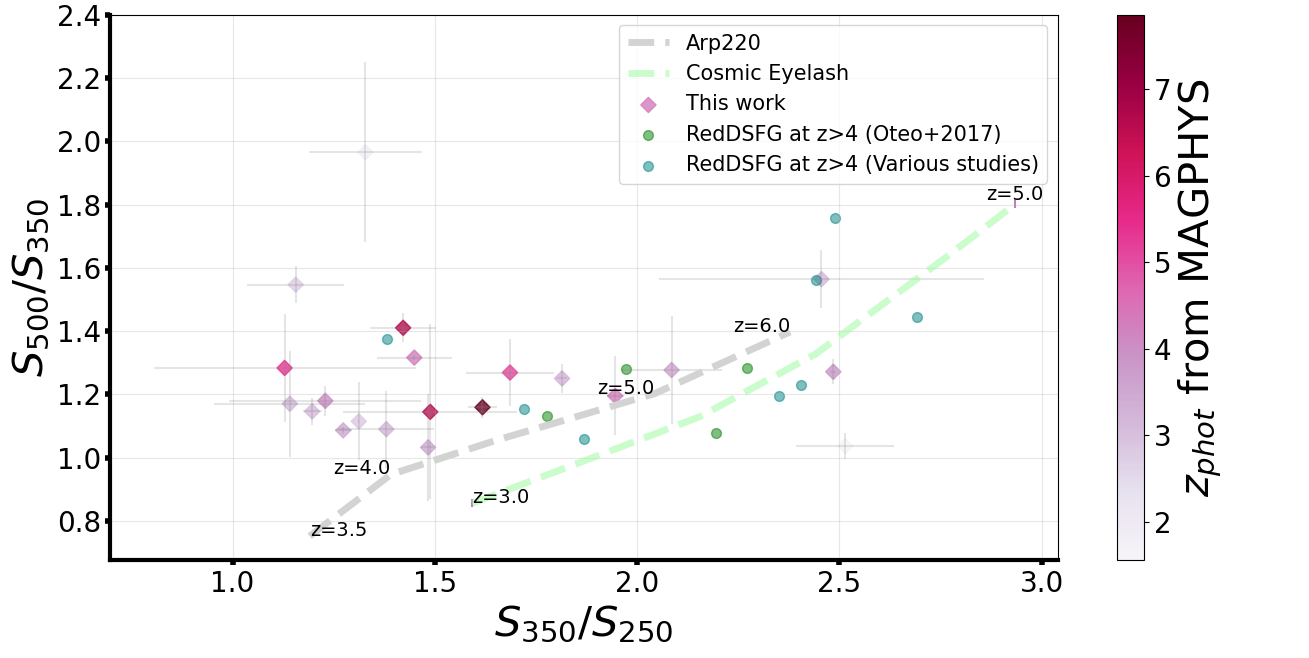}
\caption{SPIRE colour-colour diagram of our {500 $\mu$m risers}, overlaid with redshift tracks of Arp220 \citep{rag} and Cosmic Eyelash \citep{swin} and other DSFGs samples as a reference \citep{oteo}. The color scale indicates the photometric redshifts as computed by \textsc{MAGPHYS}.}
\label{fig:spirecolcol}
\end{figure*}

{Given that most of the radio-selected dust-obscured galaxies have detections in the FIR bands (145 over 272 in our sample), it is possible to exploit the trend of the flux between 250 and 500 $\mu$m to infer a lower limit for the redshift estimate. In fact, it has been shown that galaxies with $S_{250}<S_{350}<S_{500}$ are likely to sit at $z>4$ \citep{donev18,duiv,greenslade}.
We find 22 of these so-called \textit{500$\mu$m risers} in our sample (see Figure \ref{fig:histspire}). The SPIRE diagram in Figure \ref{fig:spirecolcol} displays the colours of the 500$\mu$m risers in our sample, which are compatible with other DSFGs sample at z$>$4 from the literature.
In our whole catalog, only one galaxy is above the threshold of 30 mJy defined by \citet{donev18}, with a $S_{500}\sim$34 mJy. All the others are well below this limit, suggesting a negligible impact of lensing in this analysis, as suggested by \cite{lapi14,negrello2017,donev18}.

According to the models \citep{blain}, at $S_{500}\sim$100 mJy the number counts of non-lensed galaxies drops, due to the intrinsically steep luminosity function at z$>$1.5, therefore, this flux limit is a good threshold to identify magnified sources. The {500 $\mu$m risers} in our sample have relatively low fluxes compared to those studied in the works mentioned above. This feature might not only be due to the lack of lensed sources but also to the intrinsic nature of these galaxies, which are likely{normal dust-obscured SFGs at $z>4$}, rather than extreme objects.}

\subsection{Radio}
Our analysis is based on radio detections at 3 GHz from COSMOS VLA 3 GHz Large Project, therefore all our sources have at least one radio photometric point. In order to improve our knowledge about the radio band, we also use 1.4GHz data included in the \textit{Herschel} Super-Deblended Catalog, taken from \citet{Schinnerer_2010}.

\subsection{X-ray} 
Finally, we match our sample with the Chandra catalog by \cite{civano}: 11 sources have a counterpart within 5''. However, the low resolution in these bands {makes} it challenging to understand whether the X-ray emission belongs to our radio target or a nearby COSMOS2015  source. After an accurate analysis, we conclude that for 5 of these sources, the X-ray is more likely related to the nearby NIR-bright object. For another object the association remains ambiguous, while for the remaining 6 the X-ray emission belongs most likely to our targets: two show evidence of being heavily obscured, close to the CT regime, AGN, one is an AGN with $L_X\sim10^{45}$ergs$^{-1}$, and two are obscured AGN ($n_H\gtrsim10^{23}$cm$^{-2}$).
We exclude these galaxies from the computation of the SFRD and will address their study to future works.

\section{Results from the individual SED-fitting}
\label{sec:indsed}
From the SED-fitting of the individual sources in the primary sample, we identify some of the main properties of the objects in our sample and estimate the {possible presence of obscured} AGN. 

\begin{figure}[t]
 \centering
 \includegraphics[width=0.5\textwidth]{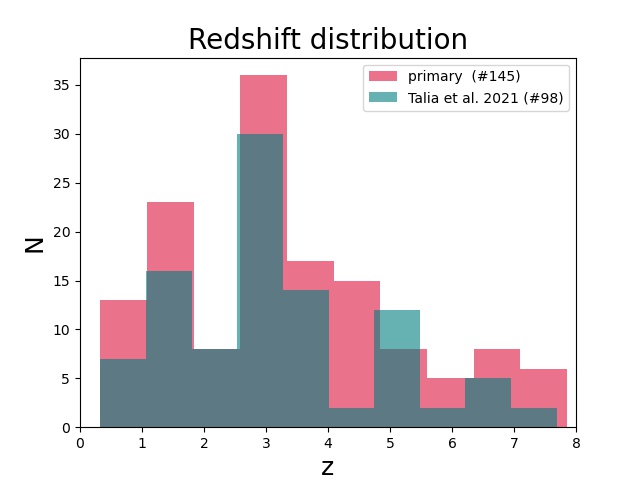} 
 \caption{Distribution of the photometric redshifts of the individual sources in this work (crimson) compared with the results by \citetalias{talia21} (teal-blue) for their similarly defined primary samples.}
 \label{fig:redshift}
\end{figure}

First of all, we show the redshift distribution in Figure \ref{fig:redshift}.
The typical error on the photometric redshift computed via SED-fitting is $\sim10\%$.   
{The usage the photometry (and upper limits) in the full available wavelength range improves the reliability of our results for the redshift estimation. As a test, we perform an SED-fitting run using only the FIR photometry, as done in other works (e.g. \citealt{Jin}): we find that the mean relative uncertainty is about 7 times higher.}

{In the following, we focus on the sources with an estimated $z_{phot} <$5, given the uncertainties in the dust properties at higher redshifts.}

The dust luminosity $L_{dust}$ has a median value over the sample of ${2.8\times10^{12}L_{\odot}}$ and it varies between $2.9\times10^{10}L_{\odot}$ and ${1.3\times10^{13}L_{\odot}}$. Thus, we conclude that most of the objects in our catalogue are ULIRGs. We also find {6} sources with $L_{dust}>10^{13}L_{\odot}$, i.e. Hyper Luminous Infrared Galaxies (HyLIRGs). These very high values for the infrared luminosity might indicate that many of these sources are going through a starburst phase. In fact, using the Kennicutt relation \citep{Kennicutt_2012} we derive a median SFR of $\sim{4.2}\times10^2M_{\odot}yr^{-1}$.

From the radio flux at 3 GHz, we derive the luminosity at 1.4 GHz for all primary sources, assuming a spectral index $\alpha$=-0.7. 
This is a fair assumption, since it corresponds to the peak of the spectral index distribution computed for the sub-sample of sources which have a counterpart at 1.4 GHz \citep[see also][]{Novak2017}.
Through the relation between infrared luminosity $L_{dust}$ and radio luminosity $L_{1.4}$, we are then able to {explore} the {possible} presence of AGN in our sample by looking at the $q_{TIR}$ parameter. Its median value $q_{TIR}\sim2.27$ is consistent with the results from \citetalias{talia21} and \cite{Novak2017} \citep[see also][]{algera2020}, suggesting no evidence for strong radio AGN emission in our sample \citep{Delhaize2017}. 

{Furthermore, one of the most interesting parameters computed by \textsc{MAGPHYS} is the fraction $f_\mu$ of dust luminosity due to diffused component of the ISM, compared to the one due to the birth clouds $(1-f_\mu)$. The younger the galaxy is, the highest the contribution of the birth clouds, since it should host large star-forming regions with newborn stars. The parameter $f_\mu$ strongly correlates with the SED shape for dusty galaxies: larger values of $f_\mu$ correspond to lower dust temperatures and SFR \citep{martis}. We can decompose the FIR luminosity into a hot (T$\sim$ 60-100 K), and warm (T$\sim$ 20 K) dust component and the ratio of these two is an efficient indicator of the dominant power in the galaxy \citep{kirc12}. The hot dust arises from the birth clouds, or possible heating from an AGN; the warm dust is associated with the diffuse ISM and is mostly composed of larger dust grains \citep{kirc20}. 

In high redshift dusty galaxies, we expect a lower  $f_\mu$ compared to the local ones since a large fraction of large grains did not have time to form at early times, and these objects are usually efficient star-formers, hosting a large number of star-forming regions. On the other hand, for galaxies without an AGN contribution, the warmer component should not be dominant. We find a median value of $f_\mu \sim 0.56$, in agreement with those reported for high redshift dusty galaxies by \cite{kirc20} and \cite{martis}. 21$\%$ of our galaxies have $f_\mu < 0.3$: this could be a hint of an extreme starburst phase or of possible AGN activity. 

In conclusion, this is further confirmation that the galaxies in our primary sample are mostly young dusty star-forming galaxies. 
We also stress that for 101 over 145 objects in our primary sample, we have photometric points in the Rayleigh-Jeans regime of the dust grey body, i.e. in the (sub)mm bands, hence a direct constraint on the $f_\mu$. However, further observations with interferometers like ALMA and NOEMA are necessary to improve our knowledge of this topic.
}

\section{Star Formation Rate Density}
\label{sec:sfrd}
\subsection{Computation of the SFRD}\label{subsec:sfrd}

\begin{figure}[t]
 \includegraphics[width= 0.45\textwidth]{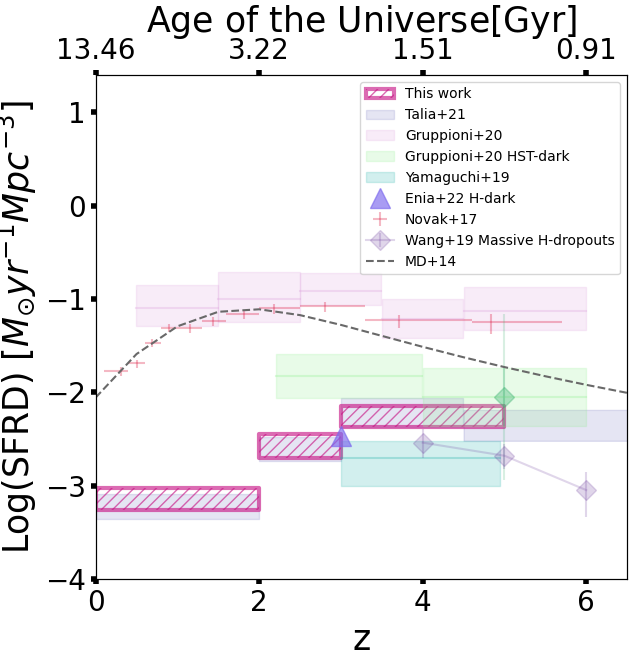} 
 \caption{Star Formation Rate Density as a function of redshift.  The crimson patterned and the violet rectangles show the confidence intervals for the results of this work and \citetalias{talia21}, where the lower boundary is the value obtained for \textit{Case-3} and the upper boundary the value obtained for \textit{Case-1}.  We  indicate  with  green, pink and turquoise the  values  found by \cite{gruppioni2020} HST-dark,  \cite{gruppioni2020} HST-bright ALMA Band-7 ALPINE detections and \cite{yamaguchi}  respectively. The violet arrow is the result by \cite{Enia} for H-dark galaxies. Red  crosses  are  the  results  by \cite{Novak2017} for COSMOS-VLA 3 GHz Large Project sources with an optical/NIR counterpart, while light green diamonds are the results by \cite{williams}, purple diamonds are by \cite{wang2019}, blue diamonds by \cite{fudamoto21} and gray circles by \cite{barrufet}. The best-fit curve by \cite{mdauadickinson}, scaled to a \citet{imf} IMF, is displayed as a dashed black line.  }
 \label{fig:SFRD}
\end{figure}

To compute the SFRD for the primary sample, we split it into four redshift bins: $0<z<2$, $2<z<3$, $3<z<5$. The {third}  bin contains about 85\% of the objects, and they are also the ones we are {more }interested in for our final purpose.

We use the $1/V_{max}$ method by \cite{vmax}. We find the maximum observable volume for each galaxy as:
\begin{equation}
 V_{max}=\sum_{z=z_{min}}^{z_{max}}[V(z+\Delta z)-V(z)]\times \frac{C_A}{C_I},
\end{equation}
where $\Delta z= 0.005$, $z_{min}$ is the lower boundary of the redshift bin and $z_{max}$ is the minimum value between the redshift upper boundary and the highest redshift at which we can detect the source, given the sensitivity of the survey.

The $C_A$ and $C_I$ constants take into
account the incompleteness due to our selection criteria.
We recall that the 272 galaxies analyzed in this work were selected from an initial sample of 476 RS-NIRdark sources only on a geometrical basis; therefore there is no point in thinking that the selected sources follow a distribution different than the one of the whole sample. 
We further divided the 272 galaxies into a \emph{primary} and \emph{secondary} samples, depending on their photometric coverage in the FIR/(sub)mm regime, and derived individual redshifts and physical properties only for the former. The galaxies in the \emph{secondary} sample might be indeed less star-forming,
or at even higher redshift, than the \emph{primary} subsample, and might
provide a lower contribution to the SFRD.
Following \citetalias{talia21}, we define the $C_I$ parameter (correction for incompleteness) as:
\begin{equation}
C_{I} = \frac{(N_{primary}+f\times N_{secondary})}{N_{primary}} \times \frac{N_{tot}}{N_{primary}+N_{secondary}}
\end{equation}
where $N_{primary}$=145 (i.e. the \emph{primary} sub-sample), $N_{secondary}$=127, $N_{tot}$=476, and $f$ is the ratio between the mean SFR$_{IR}$ of the \emph{primary} and \emph{secondary} sub-samples.

To derive the $C_I$ parameter we considered the three scenarios outlined in \citetalias{talia21}. 
In \textit{Case 1}, $f=1$ gives an upper boundary to the SFRD under the assumption that the \emph{primary} sample is representative of the whole catalog.
In \textit{Case 3} we ﬁx $f$ to the ratio between the mean FIR ﬂuxes of the
two sub-samples, derived by stacking the images of our sources in five Herschel bands (from 100 to 500 $\mu$m), SCUBA 850 $\mu$m and AzTEC 1.1mm, following \citet{bethermin2015}. In this work we find $f=0.56$, and we consider the resulting values of the SFRD as a lower boundary.

The other correction factor, $C_A$, considers the effective area (see eq.2 in \citetalias{talia21}). 

To find the SFRD in each redshift bin, we weight the SFR of every galaxy by each $V_{max}$ and sum together all the ones belonging to the same redshift range. We performed a bootstrap analysis to find the values of the SFRD and the related errors, taking into account the likelihood distribution of the redshift given by the SED-fitting code. We extracted random values of the redshift for each galaxy in the primary sample from this likelihood distribution. Thus, we inferred the value of the infrared luminosity for each of these redshifts. This way, we obtain a distribution of infrared luminosities and redshifts for every object in the sample, and we can use them to compute a distribution of SFRDs. In each bin, the final value of the SFRD is the median of the distribution. We compute the errors from the 16th and 84th percentile. 

\begin{table}[ht]
\centering
\begin{tabular}{|c|c|c|}
\hline
{\textbf{Bin}}&\textbf{SFRD ($f=0.56$)} & \textbf{SFRD ($f=1$)}  \\
 & M$_\odot$Mpc$^{-3}$yr$^{-1}$ & M$_\odot$Mpc$^{-3}$yr$^{-1}$ \\
\hline
 $0.0<z_{phot}<2.0$ & $(0.6\pm0.1)10^{-3}$ & $(0.8\pm0.1)10^{-3}$\\
 \hline
   $2.0<z_{phot}<3.0$ & $(2.4\pm0.4)10^{-3}$ & $(3.2\pm0.5)10^{-3}$ \\
   \hline
     $3.0<z_{phot}<5.0$ & $(5.0\pm0.7)10^{-3}$ & $(6.2\pm0.6)10^{-3}$\\
    \hline

\end{tabular}
\caption{Values for the SFRD found using the correction of \textit{Case 1} (column 2) and \textit{Case 3} (column 3) described in Section \ref{subsec:sfrd}.}
\label{tab:sfrd}
\end{table}

\subsection{Results}
The SFRD {as a function of redshift }is plotted in Figure \ref{fig:SFRD} and the values in each bin are reported in table \ref{tab:sfrd}.

The values at $3<z<5$ are consistent with those found by {\citetalias{talia21} for a sub-sample of our galaxies, by \cite{Enia} for a sample of radio-selected H-dark galaxies in the GOODS-N, and by \cite{gruppioni2020} for a sample of (sub)mm selected NIR-dark galaxies in the ALPINE fields}, as reported in Figure \ref{fig:SFRD}. 

In particular, \citetalias{talia21} find a SFRD equal to $(7.1\pm 1.7)\times10^{-3}M_\odot yr^{-1}$Mpc$^{-3}$ in the third redshift bin and $(5.2\pm 1.3)\times 10^{-3}M_\odot yr^{-1}$Mpc$^{-3}$ in the fourth. \cite{gruppioni2020} computed the SFRD up to $z\sim6$ for 56 sources detected in the \textit{ALMA} band 7 in the ALPINE fields. They find a SFRD of $(1.5\pm0.9)\times10^{-2} M_\odot yr^{-1}$Mpc$^{-3}$ between $z=3.5$ and $z=4.5$ and of $(0.9\pm0.7)\times 10^{-2}M_\odot yr^{-1}$Mpc$^{-3}$ between $z=4.5$ and $z=6.0$. \cite{Enia} {analysed} 17 H-dark galaxies in the GOODS-N field using the same procedure adopted in this paper and in \citetalias{talia21} {and find }the SFRD at $z\sim3$ {to be }$\sim4.510^{-3} M_\odot yr^{-1}$Mpc$^{-3}$, consistent with what we find in this work at the same redshift. 

{We stress that our estimate of the RS-NIRdark contribution to the SFRD is only a lower limit (similarly to the cases presented in \citetalias{talia21}; \citealt{gruppioni2020, Enia}) because we are not extrapolating to lower fluxes than our detection limit.}

{Comparing the SFRD from our sample of RS-NIRdark galaxies to the ones computed} through UV-selected galaxies \citep{bouwens2020,mdauadickinson}, we find that {the contribution to the SFRD of our galaxies at 3$<$z$<$5 is $\sim$10--25$\%$} of the SFRD of UV-bright ones at the same redshifts, e.g. \cite{bouwens2020}, who {studied }the SFRD in {a sample of }1362 Lyman-breakers UV-selected galaxies in the ALMA Spectroscopic Survey in the Hubble Ultra-Deep Field (ASPECS) between redshift 1.5 and 10. {They } also include the contribution to SFRD from ULIRGs. The best-fit trend of the SFRD, computed by \cite{mdauadickinson}, is {also }based mainly on UV {selected} sources. {The SFRD trend based on UV data is steeper than that found through radio data, e.g. by \cite{Novak2017}, {who} compute the SFRD for {the full} VLA-COSMOS 3 GHz Large Project \citep{smolcic2017} {sample}, complementary to the sample presented in this work.  It means that UV-based studies miss an important part of the SFRD at high redshift due to} {dusty} galaxies. {Moreover, as emerges from this analysis, the SFRD related to the more dusty objects selected in the radio band, like our RS-NIRdarks, seems important at high-redshift.} Furthermore, we might be still missing an important fraction of obscured galaxies because of technological limitations that  will be overcome thanks to new generation telescopes, such as the JWST and the Square Kilometer Array (SKA).

The value of the number density at 3$<$z$<$5, $(6.7\pm0.9)\times10^{-6}$Mpc$^{-3}$, is overall compatible with those found by \citetalias{talia21}, \cite{Enia}, \cite{Riechers_2020} and \cite{wang2019}. Semi-analytical models \citep{henriques} and hydrodynamical simulations \citep{snyder} do not predict the presence of infrared ultra-luminous dust-obscured galaxies at {high} redshift, but our results are in agreement with the number density computed by \cite{straat2014}, \cite{schreiber2018} and \cite{girelli19} for massive quiescent galaxies at $3<z<4$. This result supports the hypothesis of these galaxies being the most likely missing progenitors of massive quiescent galaxies, according to the \textit{in situ} scenario \citep{lapi18}. Future observations will be fundamental to truly understanding the role of these objects in the galaxy formation and evolution scenarios.

\section{Summary}
In this work, we estimate the contribution to SFRD given by RS-NIRdark galaxies selected from the COSMOS VLA-3 GHz Large Project. 
We find that:
\begin{itemize}
\item These objects are likely highly infrared luminous (median $L_{dust}\sim 4\times 10^{12}L_\odot$), dust-obscured galaxies at a median photometric redshift of $z\sim3.2$: they are possibly young high redshift DSFGs, as also suggested by the median value of $f_\mu\sim0.56$.
\item 76 of the 272 selected sources have a photometric redshift $>$ 3. 
\item Through the SPIRE color-color diagram, we have further proof, independent of the SED-fitting procedure, that there are  {high} redshift candidates in our sample. 
\item RS-NIRdark galaxies' contribution to the SFRD at $3<z<5$ is at least $\sim$40$\%$ of the one due to UV-selected sources, which means that the role of the \textit{dark }galaxies might be crucial to understanding the still widely unknown evolution of SFRD at z$>$3.
\end{itemize}

Our results remark on the importance of the obscured galaxies in galaxy formation and evolution and the need for a better understanding of their role in the broader cosmological framework. 
{This work} is a step towards a less biased view of the nature of these kinds of objects. 
The inclusion of more contaminated sources with respect to previous studies \citep{talia21}  allowed us to start building an improved photometric catalog and to get more information from our data on the physical properties of these galaxies. 

\section*{Acknowledgements}
We acknowledge the use of Python (v.3.8) libraries in the analysis. This work is partly based on data products from observations made with ESO Telescopes at the La Silla Paranal Observatory under ESO programme ID 179.A-2005 and on data products produced by CALET and the Cambridge Astronomy Survey Unit on behalf of the UltraVISTA consortium. M.T., A.C., A.L., M.M., A.E., F.G., C.G. and F.P. acknowledge the support from grant PRIN MIUR 2017 20173ML3WW$\_$001. M.B. ackowledges A. Battisti for adding the A3COSMOS filters to the SED-fitting code \textsc{MAGPHYS} $+$photo-z. M.B. acknowledges A. Traina and N. Borghi for all the concrete support given from the very beginning of this work and F. Gabrielli for the useful inspiring discussions.

\vspace{5mm}

\appendix
\section{Comparison with COSMOS2020}
\label{appendix}
In this work we based our selection of RS-NIRdark galaxies on the lack of a counterpart in the NIR-based COSMOS2015 catalog \citep{laigle2016}.
However, as previously stated, in our analysis we then took advantage of the UltraVISTA DR4 maps, deeper that those used to build the COSMOS2015 catalog.
During the writing of this paper, an updated version (COSMOS2020) of the multi-wavelength catalog in the COSMOS field has been published \citep{cosmos2020}, which is based on UtraVISTA DR4 maps in the NIR bands.
We matched our primary sample to the COSMOS2020 within a 0.8'' radius and we found 76 counterparts. 
In our work we did not use the results from COSMOS2020 for these sources, for consistency with the rest of our analysis: in fact the rest of our primary sample is not present in the COSMOS2020 catalog. 
Also, we decided not to treat separately these sources.

The biggest difference in the photometric depth between our catalog and the COSMOS2020 is in the MIR bands, where IRAC-1 and IRAC-2 maps \citep{moneti2022} are slightly deeper than the ones we used, as summarized in table \ref{tab_cosmos2020}. We checked the MIR fluxes of the sources that we have in common with the COSMOS2020 catalog and verified that using the deeper IRAC maps would not change our results.

\begin{table*}[ht]
\begin{center}
\caption{Comparison between the depth of the IRAC maps used in this work and the ones used in COSMOS2020.}
\begin{tabular}{|c|c|c|c|c|}
\hline
 &\textbf{IRAC-1} & \textbf{IRAC-2} & \textbf{IRAC-3} &  \textbf{IRAC-4}\\
 \hline
This work; COSMOS2015 & 25.5 & 25.5 & 23.0 & 22.9 \\
\hline
COSMOS2020 & 25.7 & 25.6 & 22.6 & 22.5 \\
\hline
\end{tabular}
\label{tab_cosmos2020}
\end{center}
\end{table*}

\begin{figure}
 \centering
 \includegraphics[width= 0.8\textwidth]{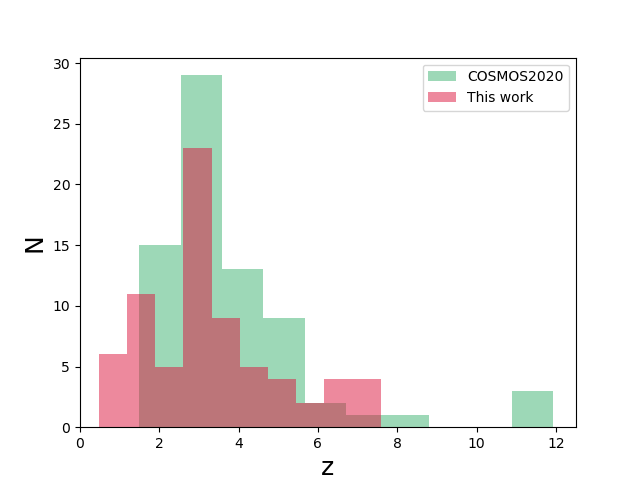} 
 \caption{Redshift distribution of the 76 sources in the primary sample with a counterpart in the COSMOS2020 catalog: in green we show the redshifts computed in this work; in red the redshifts reported in the COSMOS2020 catalog.}
 \label{fig_cosmos2020}
\end{figure}

We find also that, for the sources in common, the redshift ditributions from our own analysis and from the COSMOS2020 catalog (EAZY \citep{eazy} version) are in broad agreement (Figure \ref{fig_cosmos2020}). We stress that in our work we took advantage of the full photometric range up to radio frequencies to derive the redshifts. 

\section{SED-fitting without 350 and 500 $\mu$\lowercase{m}}
\begin{figure}
    \centering
    \includegraphics[width=0.45\textwidth]{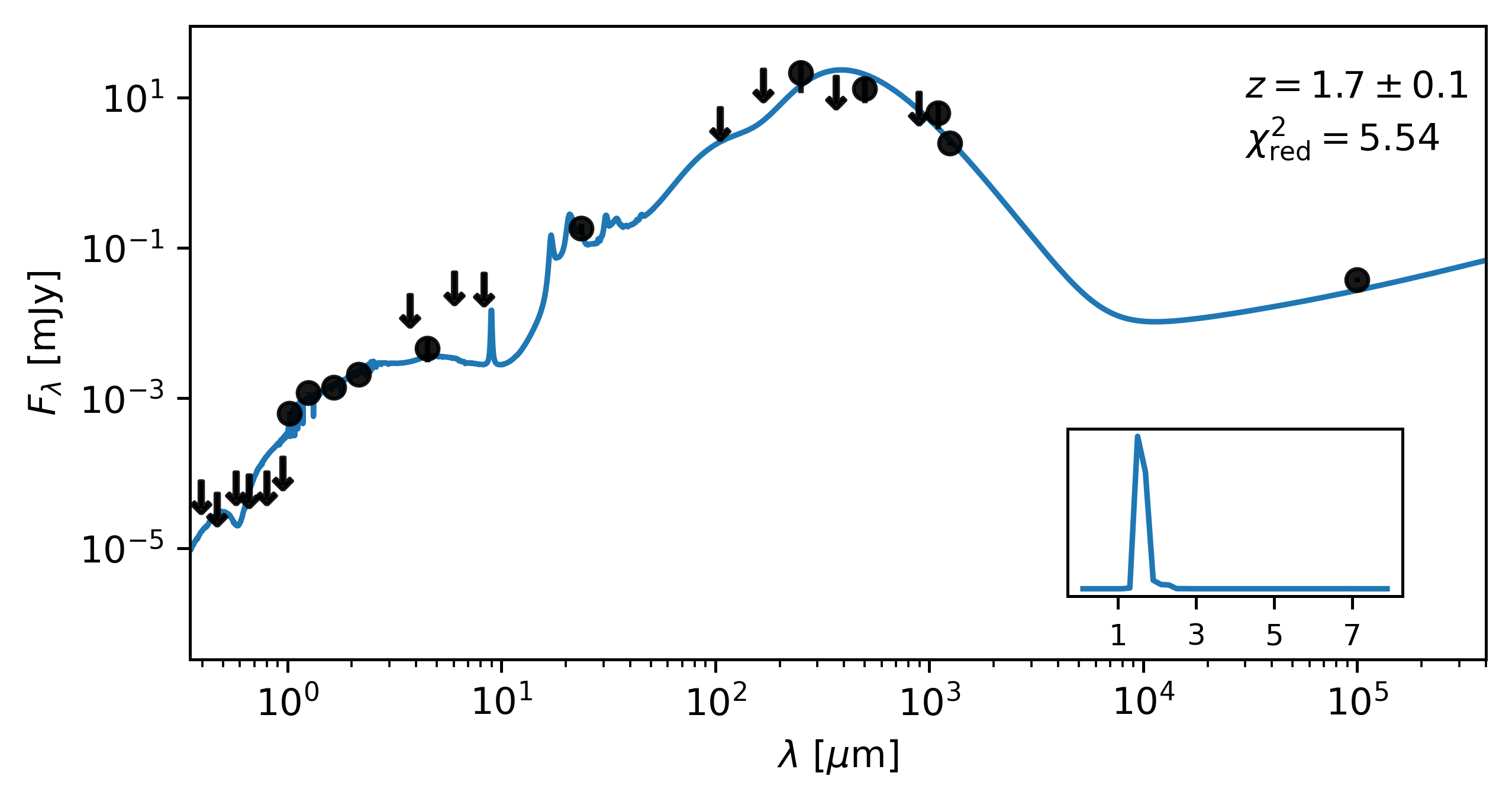}\quad    \includegraphics[width=0.45\textwidth]{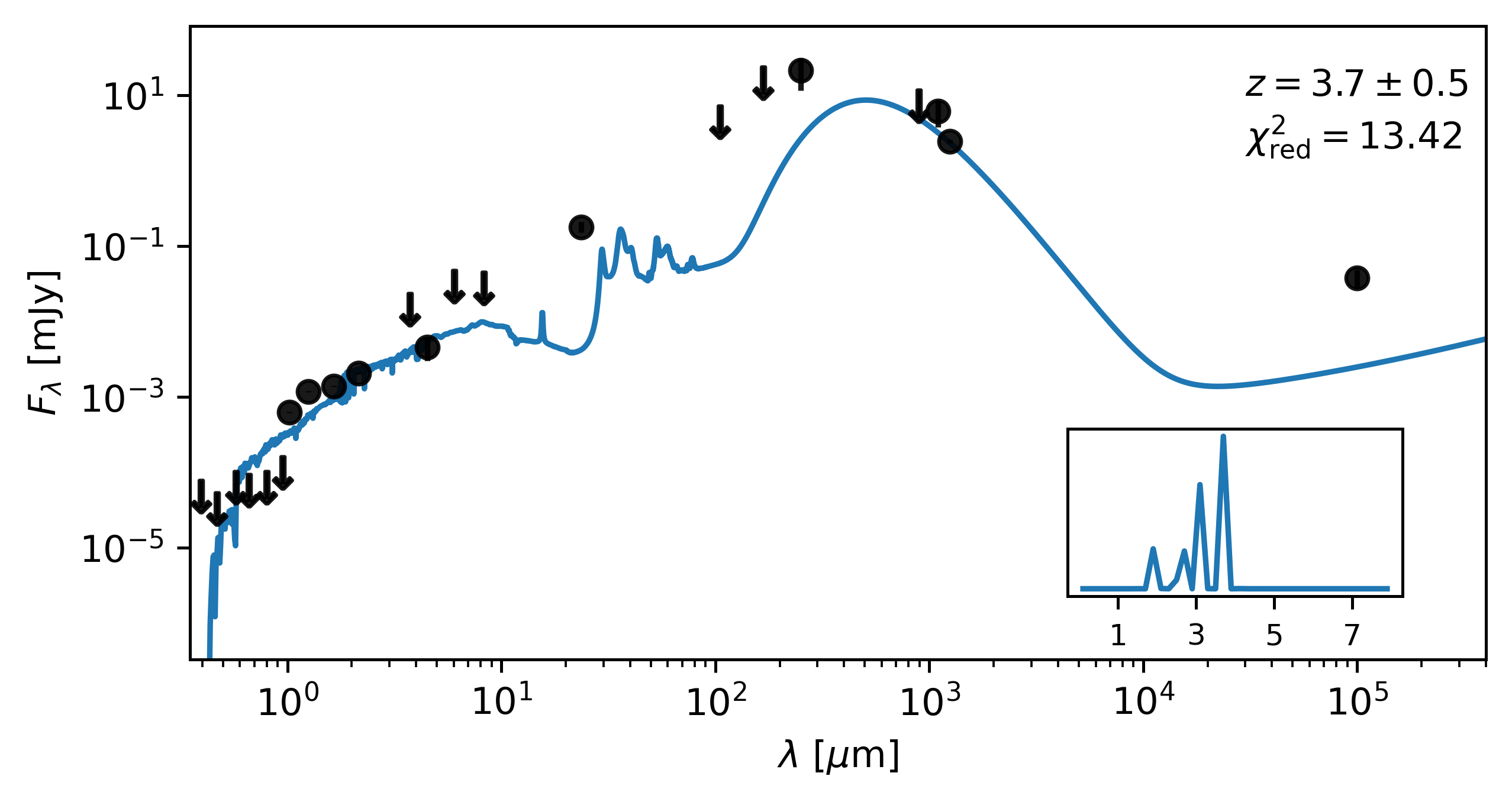}
    \caption{The best-fit SED and the probability distribution functions of the derived parameters of an example galaxy in our sample performed using the whole available photometry \textit{(left)} and without the 350 and 500 $\mu$m \textit{(right)}.}    \label{fig:mipsalmanew}
\end{figure}

\begin{figure}
    \centering
    \includegraphics[width=0.55\textwidth]{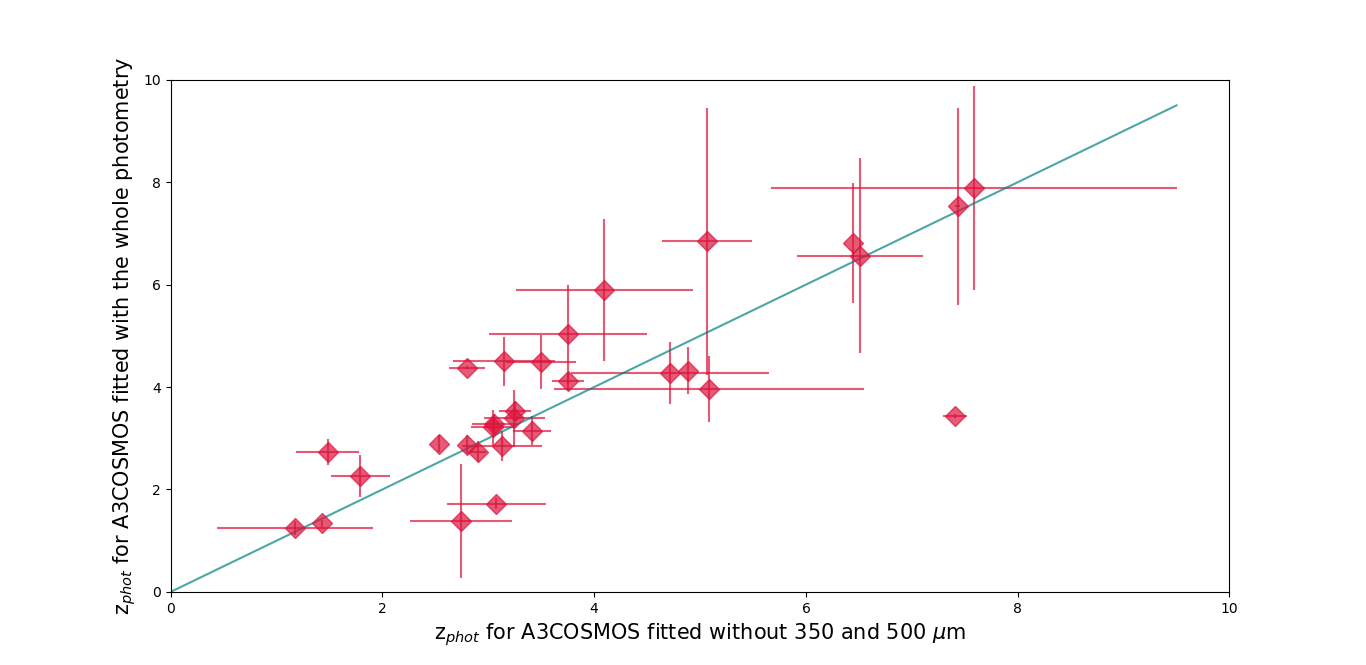}\quad  \includegraphics[width=0.35\textwidth]{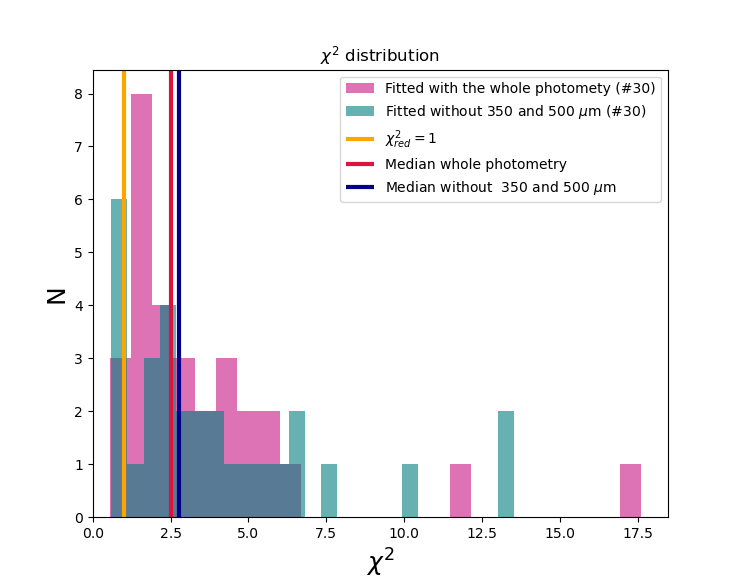}
    \caption{\textit{Left panel: }Comparison between the photometric redshifts computed with and without the 350 and 500 $\mu$m photometric data.\textit{Right panel: }Distribution of the reduced $\chi^2$ of the SED-fitting for the sources in the A3COSMOS sample obtained using the whole photometry (crimson) and without the 350 and 500 $\mu$m.}
    \label{fig:za3co}
\end{figure}

{In this appendix, we show the results when excluding from the SED-fitting of the sources with A3COSMOS counterparts the 350 and the 500 $\mu$m data points, which could be less reliable than the others because of their coarse resolution, especially with respect to the ALMA data available for this particular sub-sample. We show an example in Figure \ref{fig:mipsalmanew} and we  compare the redshifts and reduced $\chi^2$ computed in the two ways in Figure \ref{fig:za3co}.
On average, the quality of the SED-fitting stays almost the same: the median reduced $\chi^2$ for the whole photometry approach is 2.69, while for the reduced photometry it amounts to 2.87. Also, the values of the redshifts stay almost the same: the median redshift in the former case is 3.5, while in the latter it is 3.3. Moreover, the median relative error is about 0.09 for both approaches. 
}

\bibliography{sample631}{}
\bibliographystyle{aasjournal}

\end{document}